\documentclass[graybox]{svmult}

\usepackage[utf8x]{inputenc}
\usepackage{amsmath}
\usepackage{braket}
\usepackage[colorinlistoftodos]{todonotes}
\usepackage{amsfonts}
\usepackage{amssymb}
\usepackage{graphicx}
\usepackage{subfigure}
\usepackage{color}
\usepackage{ulem}
\usepackage{placeins}
\usepackage{soul,xcolor}
\usepackage{epstopdf}
\usepackage{tabu}
\usepackage{array}
\usepackage{upgreek}
\usepackage{multirow}
\usepackage{titlesec}
\usepackage{epstopdf}

\usepackage{cite}

\usepackage{mathptmx}       
\usepackage{helvet}         
\usepackage{courier}        
\usepackage{type1cm}        
\usepackage{makeidx}         
\usepackage{graphicx}        
\usepackage{multicol}        
\usepackage[bottom]{footmisc}

\makeindex             

\begin{document}

\title*{Probing signatures of beyond standard model physics through $B_s^* \rightarrow \mu^+ \mu^-$ decay}
\author{Jyoti Saini, Dinesh Kumar, Shireen Gangal and Sanjeeda Bharti Das}
\institute{Jyoti Saini \at Indian Institute of Technology Jodhpur, Jodhpur 342037, India, \email{saini.1@iitj.ac.in}
 \and Dinesh Kumar \at Department of Physics, University of Rajasthan Jaipur 302004, India, \email{dinesh@uniraj.ac.in} \and Shireen Gangal \at Center for Neutrino Physics, Department of Physics, Virginia Tech, Blacksburg, Virginia 24061, USA, \email{shireen.gangal@gmail.com}
\and Sanjeeda Bharti Das \at Department of Physics, Ramanujan Junior College, Nagaon 782001, India, \email{sanjeedabd0194@gmail.com}
}
%
%
\maketitle


\abstract{We perform  a model independent analysis to identify new physics operators which can enhance the branching ratio of $B_s^* \rightarrow \mu^+ \mu^-$ above its Standard Model (SM) prediction. We find that none of the new physics operators which provide a good fit to $b \rightarrow s \mu^+ \mu^-$ data can enhance the Br($B_s^* \rightarrow \mu^+ \mu^-$) above its SM value.}
\section{Introduction}
\label{sec:1}


Several observables related to the decays of B meson do not agree with their Standard Model (SM) predictions.   For e.g.,  the measurement of the ratios $R_{K^{(*)}}$, angular observable $P^{'}_5$ in $B \rightarrow K^* \mu^+ \mu^-$ decay in the 4.3-8.68 $q^2$-bin, branching ratio of $B_s \rightarrow \phi \mu^+ \mu^-$ do not agree with their SM value. 
  All of these discrepancies are related to the $b \rightarrow s \mu^+ \mu^-$ sector. This can be attributed to the presence of new physics (NP) in $b \rightarrow s\mu^+ \mu^-$ transition.

   In \cite{AKA121,AKA122}, new physics in $b \rightarrow s \mu^+ \mu^-$ decays were analysed  by making use of an effective Hamiltonian with all possible Lorentz structures. It was shown that any large effects in $b \rightarrow s \mu^+ \mu^-$ sector, in particular decays like $B \rightarrow K^* \mu^+ \mu^-$ and $B_s \rightarrow \phi \mu^+ \mu^-$, can only be due to new physics vector (V ) and axial-vector operators (A).  After the advent of $R_{K^{*}}$ data, several groups performed global fits  to identify the Lorentz structure of  NP \cite{Capdevila,Descotes2013,Altmannshofer382,Altmannshofer1503,
  Hurth909,Capdevila1610,Altmannshofer,AKA1703,AKA1704}. Many NP solutions, all in the form of vector V and A operators, were obtained. In order to discriminate between these NP solutions, one needs new observables. It would be desirable to have an access to observables which are theoretically clean.

The purely leptonic decay of $B_s^*$ meson is such a decay channel \cite{Grinstein1509}.
In this work we 
 perform a model independent analysis of $B^*_s \rightarrow \mu^+ \mu^-$ decay 
  to see whether Br($B^*_s \rightarrow \mu^+ \mu^-$) can discriminate between various NP solutions which provide a good fit to the $b \rightarrow s \mu^+ \mu^-$ data \cite{js}.

\section{$B_s^* \rightarrow \mu^+ \mu^-$ decay}
The effective Hamiltonian for the quark level transition $b \rightarrow s \mu^+ \mu^-$ within the SM is given by 

\begin{align} \nonumber
	\mathcal{H}_{SM} &= − \frac{4 G_F}{\sqrt{2} \pi} V_{ts}^* V_{tb} \bigg[ \sum_{i=1}^{6} C_i(\mu) O_i(\mu) + C_7 \frac{e}{16 \pi^2} [\overline{s} \sigma_{\mu \nu}(m_s P_L + m_b P_R)b]F^{\mu \nu}  \\&
	+ C_9 \frac{\alpha_{em}}{4 \pi}(\overline{s} \gamma^{\mu} P_L b)(\overline{\mu} \gamma_{\mu} \mu) + C_{10} \frac{\alpha_{em}}{4 \pi} (\overline{s} \gamma^{\mu} P_L b)(\overline{\mu} \gamma_{\mu} \gamma_5 \mu) \bigg].
\end{align}
Here $G_F$ is the Fermi constant, $V_{ij}$ are elements of the Cabibbo-Kobayashi-Maskawa (CKM) matrix and $P_{L,R} = (1 \mp \gamma^{5})/2$. The effect of the operators $O_i$, i = 1 − 6, 8 can be included in the effective Wilson Coefficients by redefining $C_7(\mu) \rightarrow C_{eff}^7(\mu, q^2)$ and $C_9(\mu) \rightarrow C_{eff}^9(\mu, q^2)$.

To study NP effects in $B^*_s \rightarrow \mu^+ \mu^-$ decay, we consider the addition of V , A, S and
P operators to the SM effective Hamiltonian of $b \rightarrow s \mu^+ \mu^-$ 
\begin{align}
	\mathcal{H}_{eff}(b \rightarrow s \mu^+ \mu^-) &= \mathcal{H}^{SM} + \mathcal{H}^{VA} + \mathcal{H}^{SP},
\end{align}
where $\mathcal{H}^{VA}$ and $\mathcal{H}^{SP}$ are as
\begin{align} \nonumber
	\mathcal{H}^{VA} &= \frac{\alpha G_F}{\sqrt{2} \pi} V_{ts}^* V_{tb} \bigg[C_9^{NP}(\overline{s} \gamma^{\mu} P_L b)(\overline{\mu} \gamma_{\mu} \mu) + C_{10}^{NP} (\overline{s} \gamma^{\mu} P_L b)(\overline{\mu} \gamma_{\mu} \gamma_{5} \mu) \\&
	+ C_9^{'NP}(\overline{s} \gamma^{\mu} P_R b)(\overline{\mu} \gamma_{\mu} \mu) + C_{10}^{'NP} (\overline{s} \gamma^{\mu} P_R b)(\overline{\mu} \gamma_{\mu} \gamma_{5} \mu) \bigg] \\ \nonumber
	\mathcal{H}^{SP} &= \frac{\alpha G_F}{\sqrt{2} \pi} V_{ts}^* V_{tb} \bigg[R_S(\overline{s} P_R b)(\overline{\mu} \mu) + R_P (\overline{s} P_R b)(\overline{\mu} \gamma_{5} \mu) \\&
	+ R_S^{'}(\overline{s} P_L b)(\overline{\mu} \mu) + R_P^{'} (\overline{s} P_L b)(\overline{\mu} \gamma_{5} \mu) \bigg]
\end{align}
where $C_9^{NP},C_{10}^{NP},C_{9}^{'NP},C_{10}^{'NP},R_S,R_P,R_S^{'},R_P^{'}$ are NP couplings.

We find that, 
$$	\langle0|\overline{s}b|B^*_s \rangle = 0,~	\langle 0|\overline{s} \gamma^5 b|B^*_s \rangle = 0.$$

Hence the Br($B^*_s \rightarrow \mu^+ \mu^-$) is not affected by NP in the form of S and P
operators.
 
The decay rate including NP V and A contribution is obtained to be,
\begin{align} \nonumber
	\Gamma(B^{*}_{s}\rightarrow\mu^{+}\mu^{-})&=\frac{G^{2}_F\alpha^{2}}{96\pi^{3}}\vert V_{tb}V^{*}_{ts}\vert^{2}f^{2}_{B^{*}_{s}}m^{2}_{B^{*}_{s}}\sqrt{m^{2}_{B^{*}_{s}}-4m^{2}_{\mu}}\bigg[~\bigg|C^{eff}_{9}(m_{B_s^*}^2) +  \\&
	2\frac{m_{b} f_{B_s^*}^T}{m_{B^{*}_{s}} f_{B_s^8}}C^{eff}_{7}(m_{B_s^*}^2) + C^{NP}_{9} 
	+C^{'NP}_{9}\bigg|^2
	+\bigg| C_{10}+ C^{NP}_{10}+C^{'NP}_{10}\bigg|^{2}~\bigg].
\end{align}

As the total decay width of $B^*_s$
meson,$\Gamma(B^{*tot}_{s})$ is not yet known precisely 
, it is assumed that $\Gamma(B^{*tot}_{s})$ is comparable to the dominant decay process $B^*_s \rightarrow B_s \gamma$ which is found to be $\Gamma (B^*_s \rightarrow B_s \gamma) = 0.10 \pm 0.05$ KeV \cite{Grinstein1509}.

\section{Methodology}
\label{subsec:3}

We perform a $\chi^2$ fit to all CP conserving data $b \rightarrow s \mu^+ \mu^-$ sector for all possible combinations of NP couplings. The observables used in the fit are listed in \cite{js}.

 The $\chi^2$ function is constructed as
\begin{align}
	\chi^2(C_i) &= (O_{th}(C_i) - O_{exp})^T \mathcal{C}^{-1} (O_{th}(C_i) - O_{exp}) .
\end{align}
The total covariance matrix $\mathcal{C}$ is obtained by adding the individual theoretical and
experimental covariance matrices.

\begin{table}
	\caption{Calculation of the branching ratios of $B_s^* \rightarrow \mu^+ \mu^-$ for various new physics scenarios. Here $\Delta \chi^2 = \chi^2_{SM} -\chi^2_{bf}$ and $\chi^2_{bf}$ is the $\chi^2$ at the best fit points.}
	\label{tab:1}
	\begin{tabular}{p{3cm}p{3cm}p{1cm}p{3cm}}
		\hline\noalign{\smallskip}
		Scenario & New physics couplings & $ \Delta \chi^2$ & Branching Ratio  \\
		\noalign{\smallskip}\svhline\noalign{\smallskip}
		$C_i=0(SM)$ & - & 0 & $(1.23 \pm 0.48) \times 10^{-11}$ \\ 
		$C_9^{NP}$ & -1.24 $\pm$ 0.18 & 43.27 & $(0.95\pm 0.48) \times 10^{-11}$ \\ 
		$C_{10}^{NP}$ & $0.91 \pm 0.19$ & 29.47 & $(1.01 \pm 0.51) \times 10^{-11}$ \\ 
		$C_9^{'}$ & 0.13 $\pm$ 0.16 & 0.66 & $(1.30\pm 0.65) \times 10^{-11}$ \\ 
		$C_{10}^{'}$ & -0.11 $\pm$ 0.13 & 0.68 & $(1.29\pm 0.65) \times 10^{-11}$ \\ 
		$C_9^{NP} = C_{10}^{NP}$ & 0.01 $\pm$ 0.18 & 0.001 & $(1.26\pm 0.64) \times 10^{-11}$ \\ 
		$C_9^{NP} = -C_{10}^{NP}$ & -0.65 $\pm$ 0.11 & 43.04 & $(0.89\pm 0.45) \times 10^{-11}$ \\ 
		$C_9^{'} = C_{10}^{'}$ & -0.04 $\pm$ 0.17 & 0.06 & $(1.26\pm 0.64) \times 10^{-11}$ \\ 
		$C_9^{'} = -C_{10}^{'}$ & 0.07 $\pm$ 0.08 & 0.81 & $(1.30\pm 0.65) \times 10^{-11}$ \\ 
		$[C_9^{NP}, C_{10}^{NP}]$ & [-1.10,0.33] & 47.33 & $(0.88\pm 0.44) \times 10^{-11}$ \\ 
		$[C_9^{'}, C_{10}^{'}]$ & [0.08,-0.07] & 0.81 & $(1.31\pm 0.66) \times 10^{-11}$ \\ 
		\begin{tabular}{@{}c@{}}$[C_9^{NP}=C_{10}^{NP}$,\\$C_9^{'}=C_{10}^{'}]$\end{tabular} & [-0.02,-0.02] & 0.07 & $(0.97\pm 0.49) \times 10^{-11}$ \\ 
		\begin{tabular} {@{}c@{}}$[C_9^{NP}=-C_{10}^{NP}$,\\
			$C_9^{'}=-C_{10}^{'}]$\end{tabular} & [-0.67,0.16] & 46.27 & $(1.00\pm 0.52) \times 10^{-11}$ \\ 
		\begin{tabular}{@{}c@{}}$[C_9^{NP},C_{10}^{NP}$,\\ $C_9^{'},C_{10}^{'}]$
		\end{tabular} & [-1.31,0.26,0.34,-0.25] & 56.04 & $(1.00\pm 0.52) \times 10^{-11}$ \\
	\noalign{\smallskip}\hline\noalign{\smallskip}
	\end{tabular}
	
	\FloatBarrier
\end{table}

\section{Results and Discussions}
\label{sec:3}
The fit results for various new physics scenarios, along with the corresponding predictions for the branching ratio of $B^*_s \rightarrow \mu^+ \mu^-$, are presented in Table 1. It is obvious from Table 1 that none of the new physics scenarios can provide large enhancement in the branching ratio of $B^*_s \rightarrow \mu^+ \mu^-$ above its SM value. In scenarios where a good fit to the data is obtained, $Br(B^*_s \rightarrow \mu^+ \mu^-)$ is seen to be suppressed as compared to the SM value. Hence, most likely, the future measurements are expected to observe $B^*_s \rightarrow \mu^+ \mu^-$ decay with a branching ratio less than its SM prediction.

\end{document}